\documentstyle[11pt, newpasp,twoside,epsf]{article}
\begin{document}

\title{On the formation of the Magellanic Stream}

\author{Chiara Mastropietro, Ben Moore, Lucio Mayer, Joachim Stadel}
\affil {Institute for Theoretical Physics, University of Z\"urich, Switzerland}
\author {James Wadsley}
\affil {Department of Physics \& Astronomy, McMaster University, Canada}

\begin{abstract}

We use high resolution N-Body/SPH simulations to study the hydro-dynamical and
gravitational interaction between the Large Magellanic Cloud and the Milky Way.  
We model the dark and hot extended halo components as well as
the stellar/gaseous disks of the two galaxies. Tidal forces
distort the LMC's disk, forcing a bar and creating a diffuse stellar halo 
and a strong warp, although very few stars are unbound from the LMC.
Ram-pressure from a low density ionised halo  is 
then sufficient to remove $1.4 \times 10^8M_\odot$ of gas from 
the LMC's disk forming a great circle trailing stream around the Galaxy.

\end{abstract}

\section{Introduction}

The Magellanic Stream (MS), a trailing filament of neutral hydrogen that
originates from the Magellanic Clouds and covers $\sim100\deg$ of the
Southern Sky, is clearly the result of an interaction between the Clouds
and the Milky Way (MW). Althought several models have been
proposed for the origin of the MS,  no single
mechanism appears able to reproduce all the key features of the stream.
In particular, tidal stripping models
(Lin \& Lyndell-Bell 1981; Gardiner \& Noguchi 1996; Weinberg 2000)
are unable to explain the lack of stars in the Stream and the gradual
decrease of the HI column density, moving from the head
(close to the Magellanic Clouds) to the tip of the stream. Early ram
pressure models (Moore \& Davis 1994; Sofue 1994) do not reproduce the
amount of gas in the Stream ($\sim 2\times 10^8 M_{\odot}$) and are unable
to explain the presence of the Leading Arm (Putman et al 1998). 
\\ The aim of this work is to
study for the first time, the
simultaneus effect of gravitational and hydrodynamical forces acting
on the Large Magellanic Cloud (LMC) as it moves in the Galactic
halo. In particular we are interested in the formation and evolution of
the Stream and in the dynamical changes in the internal structure
of the LMC due to the interaction with the MW. We neglect the effect of
the Small Magellanic Cloud owing to its small mass ($\sim$10\% of the LMC).

\section{Galactic models}
The initial conditions of the simulations are constructed using 
the technique described by Hernquist (1993). Both the LMC and
the MW are multi-component systems with a stellar and
gaseous disk, a dark halo and, eventually, a bulge.
The density profile of the NFW halo (Navarro, Frenk \& White 1997) 
is adiabatically contracted due to baryonic cooling. 
The presence of an extended hot ($\sim 10^6 $K) halo
surrounding the Galactic disk is expected by current models of
hierarchical structure formation and seems to be required in order to
explain ionisation features associated with HI structures (MS, some
high velocity clouds, Outer Spiral Arm, Complex A, Complex B,
according to Sembach et al. 2003). Constraints from dynamical
and thermal arguments fix the density of the gaseus halo in a range
between $10^{-5}$ and $10^{-4} \textrm{cm}^{-3}$ at a distance of 50
kpc from the Galactic Center. The density beyond this
radius is still unknown.  We model the hot halo of the MW with
a spherical distribution of gas that traces the dark matter profile
and is in hydrodynamical equilibrium inside the Galactic
potential. This ionised gas has a mean density of $2 \times 10^{-5}
\textrm{cm}^{-3}$ within 150 kpc and a temperature $\bar{T} \sim 10^6
$ K. The rotation curves and the main parameters of the LMC model are
shown on the left side of Figure 1. The masses and the scale lengths
are selected in order to reproduce observational constraints from Kim
et al. 1998 and van der Marel 2002, while for the MW we adopt the
favoured model of Klypin, Zhao \& Sommerville 2002.  \\

\section{Simulations}
The LMC is now at $\sim$ 50 kpc from the Galactic Center and
kinematical data imply that it is close to the perigalacticon. We wish
to simulate the past couple of orbits of the LMC whilst ending up with the
satellite in the same position and inclination as observed today. Using
several low resolution test simulations we are able to achieve an orbit
that leads to the present position and velocity of the LMC accounting
for the orbital evolution induced by the combined effect of dynamical friction
and tidal stripping (Colpi, Mayer \& Governato 1999). The result
is an orbit with the last apo/peri ratio 2.5:1, a perigalacticon
distance of $\sim$ 45 kpc and a period of 2 Gyrs.  In choosing the
initial inclination and position of the line of nodes we make the
approximation that they do not change during the interaction. In
particular we adopt the values from van der Marel 2002: respectively
$i = 34.7 \deg$ and $\Theta = 129.9 \deg$.\\ 
We follow the interaction
between the MW and the LMC for 4 Gyrs, using GASOLINE a parallel tree-SPH 
code with multistepping (Wadsley, Stadel \& Quinn 2003).
The high resolution runs have $2.46 \times 10^6$
particles, of which $3.5 \times 10^5$ are used for the disks and $5
\times 10^5$ for the hot halo of the MW. The gravitational softening
is set equal to 0.5 kpc for the dark and gaseus halos, to 0.1 kpc for
stars and gas in the disk and bulge components. 
\begin{figure}
\plottwo{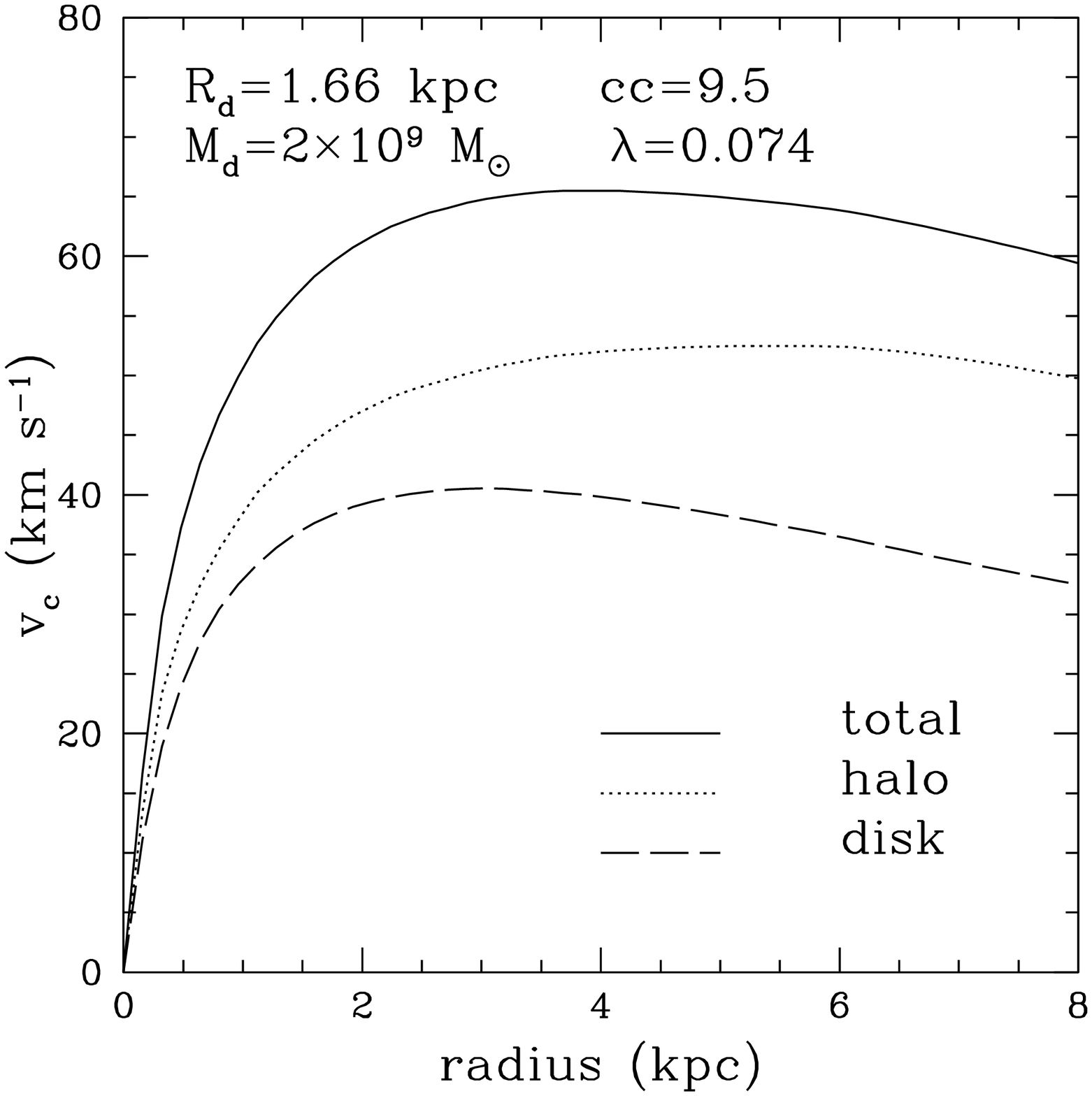}{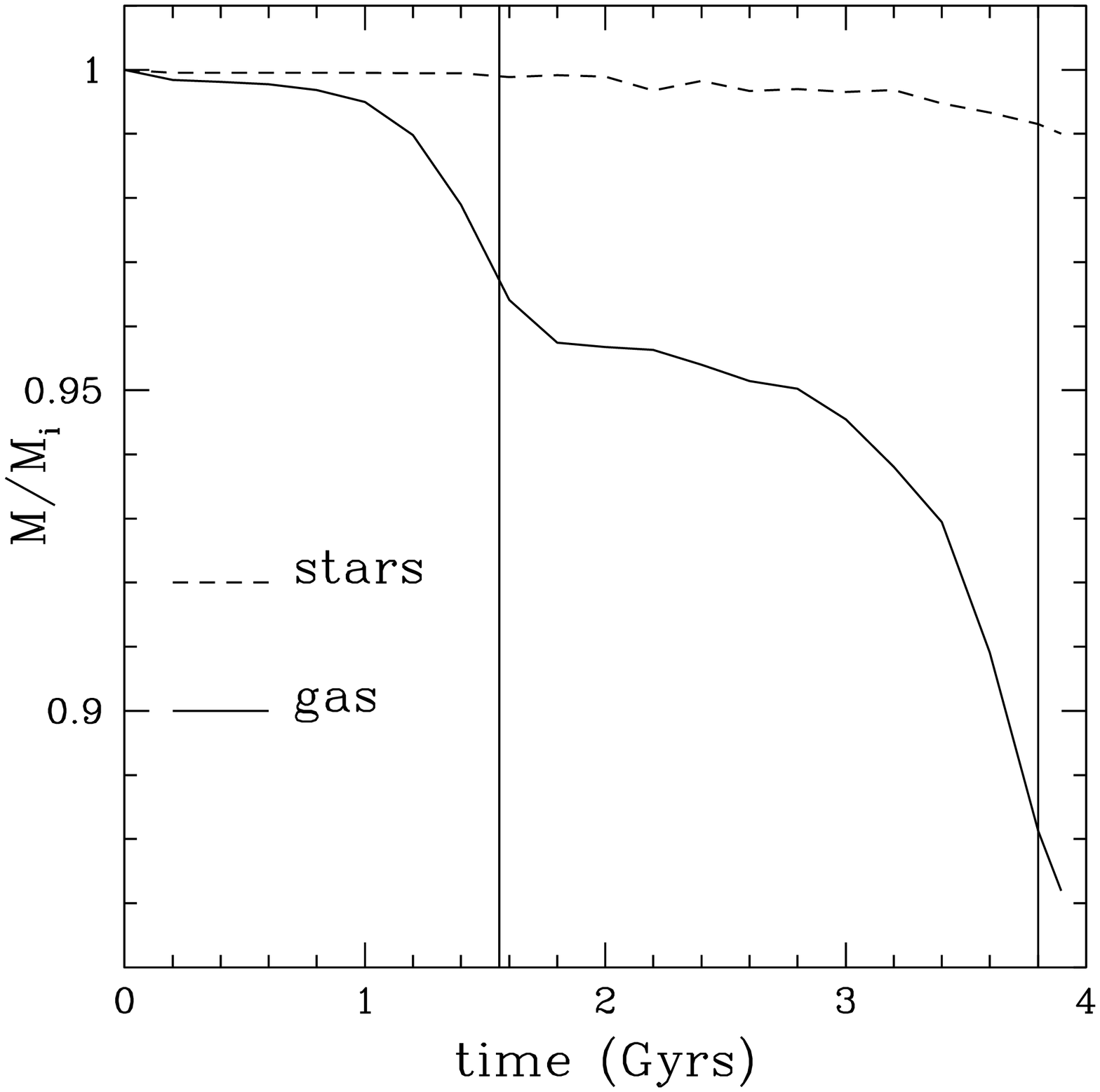}
\caption{Left: Rotation curves for the LMC model. On the top disk mass
and scale length, concentration and spin parameter are
indicated. Right: Fraction of bound stars and gas during the last 1.5
orbits. The vertical solid lines represent the apogalactica. }
\label {fig1}
\end{figure}
The right side of Figure 1 shows the loss of gas and stars (calculated
using SKID, Stadel 2001) from the LMC disk during the last 4 Gyrs. The
amount of stripped stars is negligible compared with the stripped gas. A
pure tidal stripping model would remove similarly small amounts of 
both components. The large amount of stripped gas is due primarily to
ram-pressure stripping: the rate at which 
gas is lost increases as the LMC approaches perigalacticon (the
solid vertical lines in the plot), corresponding to higher densities
of the ionised halo and to higher velocities of the satellite along
the orbit. Gas is stripped starting from the first passage at the
perigalacticon, forming a stream perpendicular to the Galactic plane
(left side of Figure 2), with a final mass of $1.4 \times 10^8
M_{\odot}$. The column density distribution gradually decreases
along the length of the stream, reaching 
$3 \times 10^{18} \textrm{cm}^{-2}$ at $100
\deg$ from the LMC. This decrease in density by nearly two orders
of magnitude along the stream is a remarkable success for the
ram-pressure scenario - tidal models generically produce streams
with surface densities that fall off much more slowly.
From Figure 2 it is evident that part of the
material stripped from the satellite during the last orbit is falling
to the Galactic center in the Northern Galactic hemisphere. Our simulations
predict that the MS forms a great circle. More speculatively, the material
stripped several Gyrs ago lies in the same place on the sky as the observed
leading arm feature.  The right side of Figure 2
shows the stripped stars: there is no well defined stellar
stream, since only $10^7M_\odot$ of stars are stripped from the disk mainly during the last
perigalacticon, when the satellite is $\approx 50$ kpc from the
Galactic center. The Galactic disk is unperturbed by the interaction, but the
LMC's disk suffers a strong warp that wraps 180 degrees around the LMC
forming a large spheroid with a tidal radius of 25 kpc.
As a consequence of
the resulting asymmetrical potential and distorted stellar disk, 
it is easier to strip more gas from the LMC's disk since the
gravitational restoring force is weaker. Thus even a low density gaseous
Galactic halo is able to remove a significant amount of gas from the LMC.
Within the first 1.6 Gyrs from the beginning of the simulation
tidal forces drive bar formation in the central 4 kpc of the satellite
galaxy. The LMC's disk becomes elongated (with an axial ratio 1:2) in a
direction close to the one of the Galactic center and perpendicular to
the Stream.

\begin{figure}
\plottwo{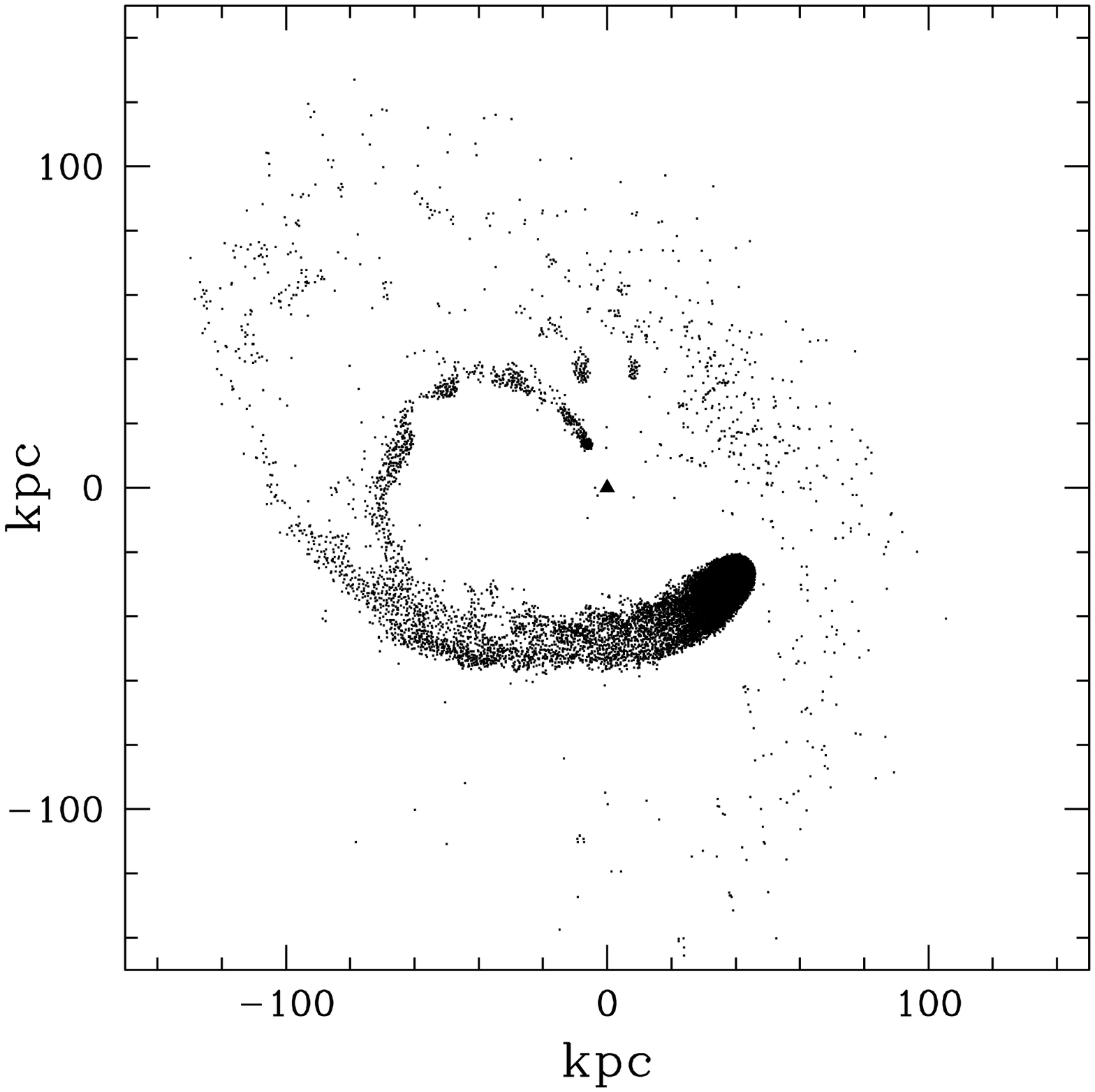}{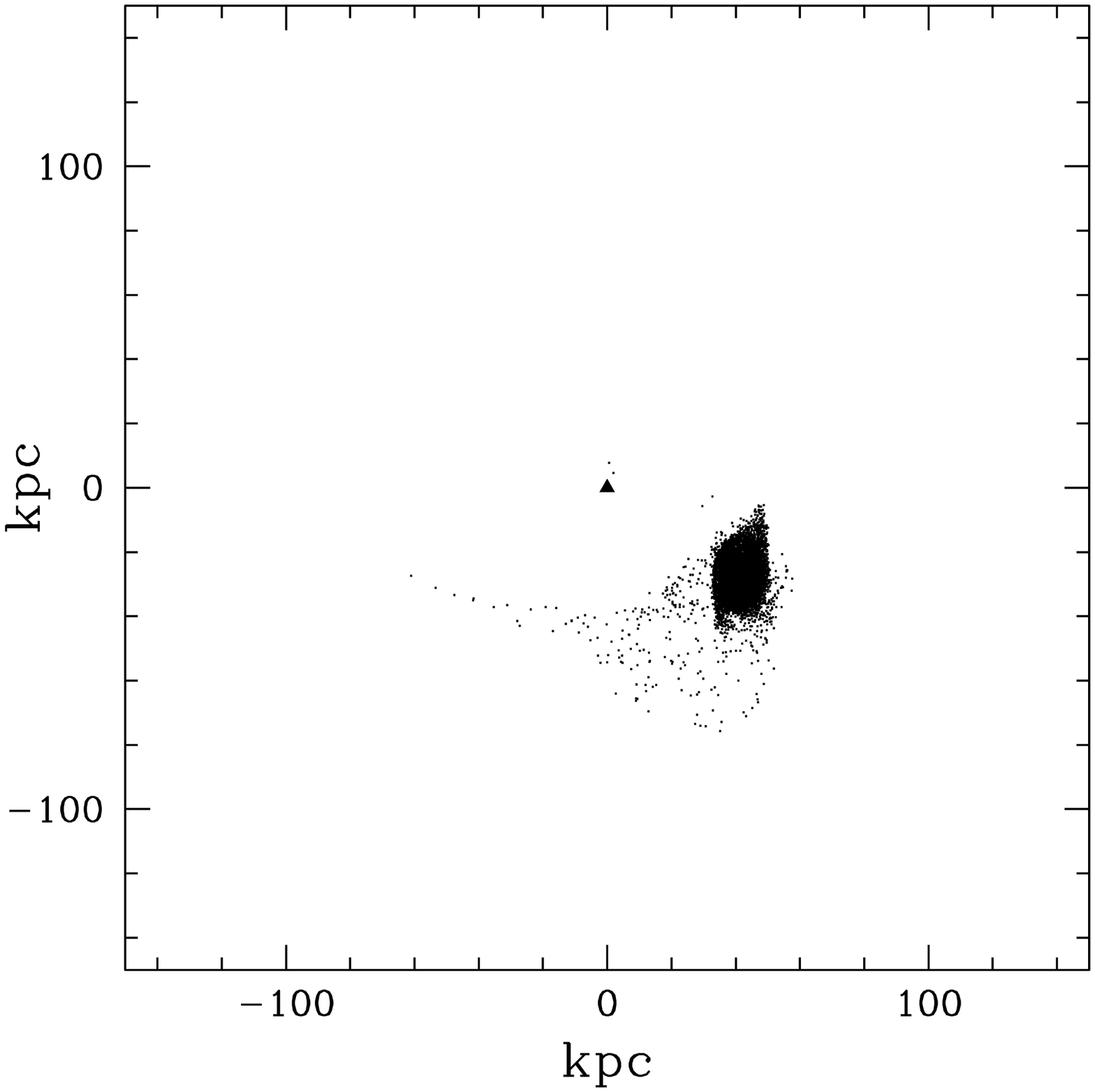}
\caption{Final distribution of gas (left) and stars (right) from the
LMC disk in a plane perpendicular to the Galactic disk.}
\label {fig2}
\end{figure}

\section{Conclusions}
1) Many of the features observed in the LMC's stellar disk can be explained
through the tidal interaction with the MW.\\
2) Tidal forces are not able to strip a significative amount of
stars from the satellite galaxy, but they produce a stellar spheroid
around the disk.\\
3) A continuous gaseus stream is produced by ram pressure stripping
from a low density gaseous Galactic halo. The morphology of the stripped
gas resembles the Magellanic Stream.\\
4) Part of the gas stripped during the first passage at the
perigalacticon is now close to the satellite. Is this gas able to
explain the Leading Arm feature?

\end{document}